%% file: main.tex
\documentclass[conference]{IEEEtran}
\IEEEoverridecommandlockouts

\usepackage{cite}
\usepackage{amsmath,amssymb,amsfonts}
\usepackage{algorithmic}
\usepackage{graphicx}
\usepackage{textcomp}
\usepackage{xcolor}

\usepackage{bm}
\usepackage{diffcoeff}
\usepackage{graphicx}
\usepackage{booktabs}
\usepackage{multirow}
\usepackage{array}
\usepackage{placeins} 
\usepackage{url}
\usepackage{breakurl}
\usepackage{xurl}
\usepackage{pgfplots}
\usepackage{filecontents}

\makeatletter
\renewcommand\section{\@startsection
    {section} 
    {1} 
    {\z@} 
    {0.5ex plus 0.5ex minus 0.5ex} 
    {0.7ex plus 0.5ex minus 0ex} 
    {\normalfont\normalsize\centering\scshape}} 
\makeatother

\makeatletter
\renewcommand\subsection{\@startsection
    {subsection} 
    {2} 
    {\z@} 
    {1.0ex plus 0.5ex minus 0.5ex} 
    {0.7ex plus .5ex minus 0ex} 
    {\normalfont\normalsize\itshape}} 
\makeatother
\def\BibTeX{{\rm B\kern-.05em{\sc i\kern-.025em b}\kern-.08em
    T\kern-.1667em\lower.7ex\hbox{E}\kern-.125emX}}
\begin{document}
\title{Robust Speech Recognition with Schrödinger Bridge-Based Speech Enhancement}
\author{
    \IEEEauthorblockN{
         Rauf Nasretdinov, Roman Korostik, Ante Juki\'{c}
    }
    \IEEEauthorblockA{
        \textit{NVIDIA}\\
        \{rnasretdinov, rkorostik, ajukic\}@nvidia.com
        \vspace{-0.5cm} 
    }
}

\maketitle

\begin{abstract}
In this work, we investigate application of generative speech enhancement to improve the robustness of ASR models in noisy and reverberant conditions.
We employ a recently-proposed speech enhancement model based on Schrödinger bridge, which has been shown to perform well compared to diffusion-based approaches.
We analyze the impact of model scaling and different sampling methods on the ASR performance.
Furthermore, we compare the considered model with predictive and diffusion-based baselines and analyze the speech recognition performance when using different pre-trained ASR models.
The proposed approach significantly reduces the word error rate, reducing it by approximately 40\% relative to the unprocessed speech signals and by approximately 8\% relative to a similarly-sized predictive approach.
\end{abstract}


\section{Introduction}
Speech signals recorded in various environments often contain adverse signal components, such as background noise or reverberation.
The goal of speech enhancement (SE) is to reduce the adverse signal components in the recorded speech and improve signal quality~\cite{hendriks2013dft, yoshioka2012making}. 
While SE is instrumental in increasing intelligibility and reducing listening fatigue, it can also be beneficial for downstream tasks, such as speaker verification, speech recognition and speech synthesis~\cite{vincent2018audio, wu2023verification, koizumi2023librittsr}.

In recent years, there has been a significant amount of research on the use of SE models as a front-end for an ASR system~\cite{narayanan2014, weninger2015speech, erdogan2015phase, aswin_2019_se_asr, kinoshita2020improving, pandey_2020_dual, ma_2021_multitaskbased, koizumi2022snri, sato2022learning, yang_2022_TimeDomainSE}. In such an approach, SE and ASR models can be trained either jointly or separately.
On the one hand, joint training of SE and ASR has been shown to benefit both tasks~\cite{weninger2015speech, erdogan2015phase, ma_2021_multitaskbased, pandey_2020_dual}. On the other hand, separate training of SE and ASR can be advantageous~\cite{kinoshita2020improving, sato2022learning, yang_2022_TimeDomainSE}.
In such a modular approach, each component may be improved separately, e.g. SE can be retrained to improve noise robustness, and ASR can be retrained to improve generalization or support multiple languages.


Traditionally, neural network-based SE systems have employed predictive approaches, learning an optimal mapping from the noisy input to optimal masks or clean speech signals~\cite{narayanan2014, williamson2015complex, xu2014regression, han2015learning, luo2018tasnet}.
Recently, several generative SE models have been proposed, aiming to improve generalization and quality of the estimated speech~\cite{lu2021study, lu2022conditional, welker22speech, richter_2023_sgmse, lemercier2023storm, scheibler2023diffusion, kamo23_interspeech, kimura_2024, lay_2023_reducing, lay2024single, jukic2024sb}.
A diffusion-based SGMSE+ model has been proposed in~\cite{welker22speech, richter_2023_sgmse}, achieving strong results in speech denoising and dereverberation.
A two-stage stochastic regeneration model (StoRM) has been proposed in~\cite{lemercier2023storm}, combining a predictive model and a diffusion-based generative model.
A generative model based on Schr\"{o}dinger bridge (SB) has been proposed in~\cite{jukic2024sb}.
As opposed to data-to-noise diffusion process, the SB describes a data-to-data process~\cite{chen_2023_sb}.
It has been shown in~\cite{jukic2024sb} that the SB-based model outperforms its diffusion-based counterparts.

Diffusion-based generative SE models have been compared with predictive SE models in several speech restoration tasks~\cite{lemercier2023analysing}. 
It has been observed that the SGMSE+ model outperforms the predictive model with the same architecture in all tasks, with the most significant improvements observed with non-additive distortion such as dereverberation and bandwidth extension.
However, the study used relatively small datasets and did not include an evaluation of ASR performance.

In this work, we analyze the benefits of generative speech enhancement on the performance of different ASR models in noisy and far-field conditions.
The contributions of this work are threefold.
First, we use a generative speech enhancement model based on Schr\"{o}dinger bridge as a front-end to ASR.
We train the model on a noisy far-field dataset and optimize the parameters of the inference process for ASR performance.
Secondly, we study the influence of different model scaling configurations on the ASR performance and compare it against a similarly-sized predictive baseline.
Thirdly, we provide detailed speech recognition results using different ASR models.
Our best SB model demonstrates strong performance, with 40.41\% relative (10.47\% absolute) improvement in word error rate (WER) compared to the input noisy speech and a 7.93\% relative (1.3\% absolute) improvement over the predictive baseline.

\section{Problem definition}
\label{sec:problem}
Consider a single speech source captured with a single distant microphone.
The time-domain signal at the microphone $\underline{\mathbf{y}} \in \mathbb{R}^N$ can be modeled as
$\underline{\mathbf{y}} = \underline{\mathbf{h}} \ast \underline{\mathbf{s}} + \underline{\mathbf{n}}$,
where $\underline{\mathbf{h}}$ is the room impulse response modeling the signal propagation in a reverberant environment, $\underline{\mathbf{s}}$ is the source speech signal, $\underline{\mathbf{n}}$ is the additive noise signal, and $\ast$ denotes convolution.
The goal of SE is to estimate the direct speech signal $\underline{\mathbf{x}}$ at the microphone from the captured microphone signal $\underline{\mathbf{y}}$. 
The models presented here operate on a short-time Fourier transform (STFT)-based time-frequency (TF) representation of the input signal with additional compression and scaling, similarly as in~\cite{richter_2023_sgmse, lemercier2023storm}.
More specifically, a TF representation $\mathbf{x} = \mathcal{A} \left( \underline{\mathbf{x}} \right) \in \mathbb{C}^D$ of the time-domain signal $\underline{\mathbf{x}}$ is obtained using the analysis transform
$\mathcal{A} \left( \underline{\mathbf{x}} \right) = b |\mathcal{F} \left( \underline{\mathbf{x}} \right) |^a \mathrm{e}^{j \angle{\mathcal{F} \left( \underline{\mathbf{x}} \right) }}$,
where $\mathcal{F}$ is the STFT operator, $|.|$ is the magnitude operator, $\angle\left(.\right)$ is the angle operator, $a \in \left( 0, 1 \right]$ is a compression coefficient, $b > 0$ is a scale coefficient, and all operations are applied element-wise.

\section{Schr\"{o}dinger bridge for speech enhancement}
\label{sec:SB}
Score-based diffusion models~\cite{song2019generative, song_2021_score, welker22speech, richter_2023_sgmse} use a continuous-time diffusion process defined by a forward stochastic differential equation (SDE)
\begin{equation}
  \label{eq:forward_sde}
  \dl{\mathbf{x}_t} = \mathbf{f} \left( \mathbf{x}_t, t\right) \dl{t} + g(t) \dl{\mathbf{w}_t} , \quad \mathbf{x}_0 = \mathbf{x}  ,
\end{equation}
where $t \in \left[ 0, T \right]$ is the current process time, $\mathbf{x}_t \in \mathbb{C}^D$ is the process state, $\mathbf{f}$ is the drift, $g$ is the diffusion coefficient, and $\mathbf{w}_t$ is the standard Wiener process. 
Schr\"{o}dinger bridge~\cite{schrodinger1932theorie, debortoli2021diffusion, chen2021likelihood, bunne2023schrodinger, chen_2023_sb} with respect to a reference path measure $p_\text{ref}$ can be defined as
\begin{equation}
  \label{eq:sb}
  \min_{p \in \mathcal{P}_{\left[0, T\right]}} D_\text{KL} \left(p, p_\mathrm{ref} \right)
  \quad
  \text{s. t.}
  \quad
  p_0 = p_x ,
  \thickspace
  p_T = p_y ,
\end{equation}
where $\mathcal{P}_{\left[0, T\right]}$ is the space of path measures on $\left[0, T\right]$, $D_\text{KL}$ is the Kullback-Leibler divergence, and $p_x$ and $p_y$ are the boundary conditions.
As shown in~\cite{chen2021likelihood,chen_2023_sb}, the SB is equivalent to a pair of forward and backward SDEs with additional drift terms.
Solving the SB is in general intractable, but closed-form solutions can be derived in special cases, such as with Gaussian boundary conditions~\cite{bunne2023schrodinger, chen_2023_sb}.
A tractable form of SB can be derived for paired data assuming Gaussian boundary conditions $p_0 = \mathcal{N}_\mathbb{C} \left( \mathbf{x}, \epsilon_0^2 \mathbf{I} \right)$ and $p_T = \mathcal{N}_\mathbb{C} \left( \mathbf{y}, \epsilon_T^2 \mathbf{I} \right)$ with $\epsilon_T = \mathrm{e}^{\int_0^T f(\tau) \dl{\tau}} \epsilon_0$, $\epsilon_0 \to 0$, and a linear drift $\mathbf{f} \left( \mathbf{x}_t, t \right) = f(t) \mathbf{x}_t$~\cite{chen_2023_sb}.
The drift scale $f(t)$ and diffusion $g(t)$ define the noise schedule for the process, and different schedules are used in the literature~\cite{chen_2023_sb, jukic2024sb}.
Here we employ the commonly-used variance-exploding noise schedule defined as $f(t) = 0$ and $g(t) = \sqrt{c} k^t$, parametrized with $k, c > 0$ as in~\cite{lay_2023_reducing}.
\subsection{Training}
\label{subsec:model_training}
The backbone neural model $d_\theta$ is trained using the data prediction loss~\cite{chen_2023_sb}, aiming to predict the target $\mathbf{x}$ from the provided inputs.
As in~\cite{jukic2024sb}, we use an auxiliary $\ell_1$-norm time-domain loss to improve the estimate of the model, resulting in the following training objective
\begin{equation}
    \label{eq:data_prediction_loss}
    \resizebox{0.9\columnwidth}{!}{$\displaystyle{
        \min_\theta \;
        \mathcal{E}_{\left( \mathbf{x}, \mathbf{y} \right), t, \mathbf{z}} \frac{1}{D}
        \lVert \hat{\mathbf{x}}_\theta(t) - \mathbf{x} \rVert_2^2
        + \lambda \| \mathcal{A}^{-1} \left( \hat{\mathbf{x}}_\theta(t) \right) - \underline{\mathbf{x}} \|_1
        }$
    }
\end{equation}
where $\mathcal{E}$ denotes the mathematical expectation, $\hat{\mathbf{x}}_\theta(t) = d_\theta(\mathbf{x}_t, \mathbf{y}, t)$ denotes the output of the backbone neural network, $\mathbf{x}_t$ is a sample from the marginal distribution $p_t$, $\mathcal{A}^{-1} \left( \hat{\mathbf{x}}_\theta(t) \right)$ is the estimated time-domain signal, $\mathbf{z} \sim \mathcal{N}\left( \mathbf{0}, \mathbf{I} \right)$ and $\lambda > 0$ is a tradeoff parameter.
\subsection{Inference}
\label{subsec:inference}
Given an initial value $\mathbf{x}_\tau$ at time $\tau>0$, the bridge SDE solution at time $t < \tau$ can be obtained as~\cite{chen_2023_sb}
\begin{equation}
    \label{eq:sde_sampler}
    \resizebox{0.9\columnwidth}{!}{$\displaystyle{
        \mathbf{x}_t = \frac{\alpha_t \sigma_t^2}{\alpha_\tau \sigma_\tau^2} \mathbf{x}_\tau
        + \alpha_t \left( 1 - \frac{\sigma_t^2}{\sigma_\tau^2} \right) \hat{\mathbf{x}}_\theta(\tau) 
        + \alpha_t \sigma_t \sqrt{ 1 - \frac{\sigma_t^2}{\sigma_\tau^2} } \mathbf{z} ,
        }$
    }
\end{equation}
where $\mathbf{z} \sim \mathcal{N}\left( \mathbf{0}, \mathbf{I} \right)$ and $\alpha_t$ and $\sigma_t$ are computed using $f(t)$ and $g(t)$~\cite{chen_2023_sb, jukic2024sb}.
Similarly, using probability flow ordinary differential equation (ODE) formulation, the bridge ODE solution can be obtained as~\cite{chen_2023_sb}
\begin{equation}
    \label{eq:ode_sampler}
    \resizebox{0.9\columnwidth}{!}{$\displaystyle{
        \mathbf{x}_t = \frac{\alpha_t \sigma_t \bar{\sigma}_t}{\alpha_\tau \sigma_\tau \bar{\sigma}_\tau} \mathbf{x}_\tau + \frac{\alpha_t}{\sigma_T^2} \left( \bar{\sigma}_t^2 - \frac{\bar{\sigma}_\tau \sigma_t \bar{\sigma}_t}{\sigma_\tau} \right) \hat{\mathbf{x}}_\theta(\tau)
        + \frac{\alpha_t}{\alpha_T \sigma_T^2} \left( \sigma_t^2 - \frac{\sigma_\tau \sigma_t \bar{\sigma}_t}{\bar{\sigma}_\tau} \right) \mathbf{y} ,
        }$
    }
\end{equation}
where $\bar{\sigma}_t$ is computed using $f(t)$ and $g(t)$~\cite{chen_2023_sb, jukic2024sb}.
Both the SDE sampler in~\eqref{eq:sde_sampler} and the ODE sampler in~\eqref{eq:ode_sampler} start from $\mathbf{x}_T = \mathbf{y}$ and iterate through a number of steps, resulting in an estimate $\hat{\mathbf{x}} = \mathbf{x}_0$.
The time-domain output signal is obtained by inverting the analysis transform as $\underline{\hat{\mathbf{x}}} = \mathcal{A}^{-1} \left( \hat{\mathbf{x}} \right)$.

\subsection{Neural model}
\label{subsec:neural_model}
As a backbone neural network for the SB model, we use the noise-conditional score network (NCSN++) as in~\cite{song_2021_score, richter_2023_sgmse, lemercier2023storm, lemercier2023analysing}. 
The model incorporates U-Net structure with four downsampling and upsampling layers following the architecture described in~\cite{lemercier2023storm}.
However, we did not use any attention layers. 
In each resolution layer, a downsampled spectrogram transformed by a two-dimensional convolutional layer is provided as a residual connection.
These resolution layers consisted of BigGAN residual blocks~\cite{brock_2018_biggan}, where channels are upconverted and downconverted within each block using two-dimensional convolutional layers.
In our study, we experimented with different number of channels at every resolution layer and different number of residual blocks within all hidden blocks.

\input{tables/table_models}

\input{tables/table_comparison}

\section{Experiments}
\label{sec:experiments}
\subsection{Datasets}
\label{subsec:datasets}

The generated training set consisted of approximately 200 hours of audio and the validation and test sets consisted of approximately 10 hours of audio, at a sample rate of 16~kHz.
We simulated 10k rooms for the training set, and 200 rooms each for the development and test sets with varying sizes, reverberation times in $[0.1, 0.5]$\,s and microphone placement as in~\cite{jukic2023waspaa}.
Clean subsets from LibriSpeech~\cite{panayotov2015librispeech} were used as the clean speech signals.
CHiME-3~\cite{barker2017third} and DNS Challenge datasets~\cite{dubey2022icassp} were used as the noise signals.
The reverberant signal-to-noise ratio (RSNR) for noisy signals was uniformly sampled from $[-5, 20]$\,dB.
In all experiments, the noisy input $\underline{\mathbf{y}}$ is the signal from the microphone closest to the speech source, and the target speech signal $\underline{\mathbf{x}}$ is the direct speech component at the same microphone.
\subsection{Experimental setup}
We employed STFT with window size of 510 samples ($\approx$32ms) and hop size of 128 samples (8ms) with a Hann window and $a = 0.5$ and $b = 0.33$ as in~\cite{welker22speech, lemercier2023storm}.
We experimented with alternative compression parameters, but did not observe any improvements.
The loss in~\eqref{eq:data_prediction_loss} used $\lambda = 10^{-3}$, as in~\cite{jukic2024sb}.
Training without the time-domain loss resulted in a small performance degradation.
The variance exploding noise schedule was parameterized with $k = 2.6$ and $c = 0.40$, as in~\cite{lay_2023_reducing, jukic2024sb}. 
Training used randomly-selected audio segments of 256 STFT frames with input and target signals normalized to the maximum amplitude of the input signal.
The global batch size was set to $\left\lbrace 64, 32, 16 \right\rbrace$ for models with $\left\lbrace 25, 50, 100 \right\rbrace$ million parameters respectively, and Adam optimizer was used with a learning rate of $10^{-4}$~\cite{lemercier2023storm}.

As in~\cite{lemercier2023analysing}, a predictive model is used as a baseline, using the same backbone neural network as the corresponding SB model. 
In this model, the neural network was trained to directly estimate the clean speech coefficients $\hat{\mathbf{x}}$ from the noisy input $\mathbf{y}$, similarly as in~\cite{lemercier2023storm}.
The conditioning layers are removed in the predictive model as in~\cite{lemercier2023storm}, without a large influence on the number of parameters (5\% decrease).
As an additional baseline, we used the stochastic regeneration model denoted StoRM~\cite{lemercier2023storm} with the same backbone neural network.
All models were trained on eight NVIDIA V100 GPUs for a maximum of 200 epochs. 
An exponential moving average (EMA) of the weights with 0.999 decay was used~\cite{lemercier2023storm}, and the best EMA checkpoint was selected based on the Perceptual Evaluation of Speech Quality (PESQ) metric value of 50 validation examples, similarly as in~\cite{richter_2023_sgmse,lemercier2023storm, jukic2024sb}. 
Inference utilized a uniform time grid with 50 time steps, unless otherwise specified. 
Predictive model was trained with mean squared-error loss function in time-domain.
StoRM consisted of SGMSE+ model trained using score matching and predictive NSCN++ module.
Checkpoint was chosen the same way as for the SB model.
All models were implemented using NVIDIA's NeMo toolkit~\cite{nvidia_nemo_toolkit}.

ASR performance is evaluated using three different ASR models:
(\textit{i}) Fast Conformer Transducer Large model with 120M parameters~\cite{fastconformer}, trained on 25k hours of English speech~\cite{fastconformer_transducer_large}
(\textit{ii}) Parakeet RNNT~\cite{parakeet_rnnt_1.1B} and (\textit{iii}) Parakeet CTC~\cite{parakeet_ctc_1.1B} with 1.1B parameters trained on 64k hours of English speech~\cite{parakeet_blog}.
Unless stated otherwise, the ASR results are obtained using Fast Conformer Transducer Large.

As speech enhancement metrics we measured perceptual evaluation of speech quality (PESQ)~\cite{rix_2001_pesq} and scale-invariant signal-to-distortion ratio (SI-SDR)~\cite{leroux_2019_sdr}. 
\input{tables/table_asr_models}
\input{figures/number_steps}

\input{figures/noise_levels_pred}
\subsection{Results}
\subsubsection{Architecture search}
\label{sssec:architecture}
In this ablation study, we investigated the impact of the model architecture in terms of number of parameters, residual blocks and channels within the NSCN++ model architecture.
The results obtained with various model configurations are shown in Table~\ref{table:models}.
For simplicity, here we consider only the results with the SDE sampler with 50 sampling steps.
The first row shows the results obtained by a 25M baseline model with three residual blocks and channels [${N}_1$, ${N}_2$, ${N}_3$, ${N}_4$] are set to [128, 128, 128, 256].
At first, we increased the model's size to 50M by either increasing the number of residual blocks (cf. config 3 in Table~\ref{table:models}) or adjusting the channels (cf. config 5 in Table~\ref{table:models}).
Interestingly, increasing the number of blocks showed notably superior results. 
Increasing the number of blocks to nine (cf. config 7 in Table~\ref{table:models}) or a combination of adjusting both channels and number of blocks (cf. config 9 in Table~\ref{table:models}) resulted in a degradation in the ASR performance. 
Notably, in terms of SE metrics, the model with six residual blocks did not show the best performance in either PESQ or SI-SDR.
This shows that ASR performance does not necessarily correlate with SE metrics.
Therefore, we used the 50M model with six residual blocks in all subsequent experiments, since it provided the best ASR performance.
\subsubsection{Sampler}
\label{sssec:sampler}
In this ablation study, we investigated the influence of the sampler for the reverse process on the ASR performance, and results with either the SDE sampler in~\eqref{eq:sde_sampler} or the ODE sampler in~\eqref{eq:ode_sampler} with 50 sampling steps are shown in Table~\ref{table:models}.
The results demonstrate a superior efficacy of the ODE sampler across almost all model configurations, with the exception of the 25M model. 
The improved performance of the ODE sampler is due to its robustness against the hallucination problem.
Generative models can produce speech-like sounds during very noisy periods, especially at low RSNR levels, leading to high rates of insertions and substitutions.
As shown in Table~\ref{table:models}, the ODE sampler significantly reduces insertions and substitutions compared to SDE sampler, while simultaneously reducing the WER, indicating its enhanced stability and robustness.
Therefore, we used the ODE sampler for the subsequent experiments.
\subsubsection{Number of sampling steps}
\label{sssec:steps}
In this ablation study, we investigated the influence of the number of sampling steps on the ASR performance, and results in terms of WER obtained are shown in Figure~\ref{fig:num_steps}.
The results indicate that the ASR performance can be improved by selecting an appropriate number of steps.
In general, WER improves significantly when the number of steps is increased from five to ten.
However, WER begins to slightly deteriorate when the number of steps exceeds 15.
Interestingly, SE performance shows similar pattern: the highest SI-SDR value is obtained at 15-20 reverse steps, but it starts to slightly degrade as this parameter increases further.
Since the difference in ASR performance between ten and fifteen steps is not substantial, ten steps is the optimal value as it is approximately 1.5 times faster.
Therefore, we used ten reverse steps in all subsequent experiments.
\subsubsection{Comparison with baseline models}
\label{sssec:comparisons}
In this study, we compared the selected SB model with baseline models.
Comparison of our SB-based model, StoRM and a predictive model in terms of ASR performance is provided in Table~\ref{table:compare}.
All models shared the same NCSN++ architecture, as configured in the third row in Table~\ref{table:models}.
The table shows that the selected SB model resulted in 1.3\% (7.93\% relative) improvement in terms of WER compared to the predictive model.
Overall, the selected SB model resulted in 10.47\% (40.41\% relative) improvement in terms of WER compared to the unprocessed speech.
When compared to StoRM model, the proposed SB model showed 4.92\% (24.16 \% relative) WER improvement. 
We tried to train StoRM with implementation provided by the authors of the model but this led to slightly worse results than using our implementation. 
\subsubsection{Evaluation across noise levels}
\label{sssec:noise}
In this study, we investigated the performance of the selected models across different noise levels and results in terms of ASR performance are shown in Figure~\ref{fig:noise_levels}.
The use of the SB model improves WER across all noise levels.
The biggest improvements compared to unprocessed speech in relative WER reduction, around 50\%, are observed at 0dB and 5dB RSNR.
In these conditions noise is high enough to degrade ASR performance significantly but still low enough for the SE model to effectively reconstruct speech and improve ASR performance.
The relative WER improvement at higher RSNRs, 10dB and 15dB, is approximately 27\%, which is slightly lower but still significant.
Even at 20dB RSNR, where the input speech is nearly clean, there is still a 2.5\% relative WER improvement.
Besides, Figure~\ref{fig:noise_levels} illustrates that the predictive model shows ASR degradation in low-noise scenarios compared to unprocessed speech.
This may be due to artifacts introduced into the processed signal that are typical of predictive models.
The SB model does not have such a disadvantage, showing improved performance even at low-noise levels.
\subsubsection{Evaluation using different ASR models}
\label{sssec:asr_models}
In this study, we investigated the use of the SB model as a front-end to different ASR models and results for Fast Conformer Transducer Large, Parakeet RNNT 1.1B and Parakeet CTC 1.1B, are shown in Table~\ref{table:asr_models}.
Using the SB model for Fast Conformer Transducer Large noticeably reduced the percentage of deletions by a factor of four, resulting in 40\% relative WER improvement.
Similarly, for Parakeet RNNT, utilizing the SB model as a preprocessing step significantly improved relative WER by 35\%, achieving the best WER on our test set of approximately 13\%.
Interestingly, although Parakeet RNNT with the SB front-end outperformed Fast Conformer Transducer Large with the SB front-end, the latter performed better than standalone Parakeet RNNT. 
Furthermore, Parakeet CTC performed worse than RNNT. Initially, both models had the same WER for unprocessed signals, but RNNT had more deletions (15.88\%) and fewer substitutions (3.96\%) compared to CTC (12.39\% and 7.63\%).
After SB enhancement, deletions dropped to around 4\% for both models, but CTC had more substitutions (10.09\% vs. 7.73\% for RNNT).
In summary, the SB front-end resulted in significant improvements for all considered ASR models.
\section{Conclusion}
\label{sec:conclusion}
In this paper we analyzed speech recognition improvement by integrating generative speech enhancement model based on Schr\"{o}dinger bridge as a pre-processing step to ASR. 
The optimal model configuration was selected based on the provided ablation studies.
Trained on a speech dataset with noise and reverberation, the best SB model significantly reduced the WER, achieving a 40\% relative improvement compared to unprocessed speech and an 8\% relative improvement compared to a predictive model of the same size.
Additionally, we demonstrated that the SB model enhances ASR performance across a variety of speech recognition models of different sizes and configurations.
\FloatBarrier
\cleardoublepage
%

\input{main.bbl}
%
\end{document}

%% file: tables/table_models.tex
\begin{table*}
\caption{Performance of different backbone architecture configurations in terms of ASR and SE metrics. }
\vspace{-0.5cm}
\label{table:models}
\begin{center}
  \setlength\tabcolsep{3.0pt} 
  \renewcommand{\arraystretch}{0.7}
  \resizebox{0.9\textwidth}{!}{
    \begin{tabular}{ccccccccccc}
      \toprule
      \multirow{2}{*}{Configuration} & \multirow{2}{*}{Parameters/M} & \multirow{2}{*}{Channels} & \multirow{2}{*}{Residual blocks} & \multirow{2}{*}{Sampler} & \multicolumn{4}{c}{ASR Metrics} & \multicolumn{2}{c}{SE Metrics} \\ \cmidrule(lr){6-9} \cmidrule(lr){10-11}
      & & & & & WER/\%~$\downarrow$ & INS/\%~$\downarrow$ & DEL/\%~$\downarrow$ & SUB/\%~$\downarrow$ & PESQ~$\uparrow$ & SI-SDR/dB~$\uparrow$ \\ \midrule
      1  &     25  &    [128,128,128,256] &    3 &   SDE &   22.90 & 1.16 & 8.63  & 13.11 &   1.85 &   3.35 \\ 
      2  &     25  &    [128,128,128,256] &    3 &   ODE &   23.07 & 0.41 & 12.74 & 9.92  &   1.44 &   4.25 \\ \midrule
      3  &     50  &    [128,128,128,256] &    6 &   SDE &   19.25  & 1.42 & 3.52  & 14.31  &   1.79 &   4.07 \\ 
      4  &     50  &    [128,128,128,256] &    6 &   ODE &   16.73 & 0.47 & 7.63  & 8.63  &   1.53 &   5.60 \\  \midrule
      5  &     50  &    [256,256,256,512] &    3 &   SDE &   20.78 & 0.98 & 7.64  & 12.16 &   1.90 &   5.11 \\ 
      6  &     50  &    [256,256,256,512] &    3 &   ODE &   18.57 & 0.38 & 10.18 & 8.00  &   1.52 &   5.23 \\ \midrule
      7  &     100 &    [128,128,128,256] &    9 &   SDE &   20.36 & 1.13 & 7.49  & 11.73 &   1.89 &   5.24 \\ 
      8  &     100 &    [128,128,128,256] &    9 &   ODE &   18.28 & 0.39 & 9.85  & 8.05  &   1.52 &   5.49 \\ \midrule
      9  &     100 &    [256,256,256,512] &    6 &   SDE &   20.66 & 1.08 & 6.77  & 12.80  &   1.78 &   4.82 \\ 
      10 &     100 &    [256,256,256,512] &    6 &   ODE &   16.39 & 0.40 & 8.03  & 7.96  &   1.53 &   6.73 \\ \bottomrule
    \end{tabular}
    }
\end{center}
\end{table*}

%% file: tables/table_comparison.tex
\begin{table}
\caption{Comparison of the selected SB model with baseline models.}
\vspace{-0.45cm}
\label{table:compare}
\begin{center}
  \setlength\tabcolsep{5.0pt} 
  \renewcommand{\arraystretch}{0.7}
  \resizebox{.85\columnwidth}{!}{
    \begin{tabular}{lrrrr}
      \toprule
      Signal      & WER/\%~$\downarrow$ & INS/\%~$\downarrow$ & DEL/\%~$\downarrow$ & SUB/\%~$\downarrow$ \\ \midrule
      Clean       &  2.07 & 0.15 &  0.29 & 1.61 \\ 
      Unprocessed & 25.91 & 0.27 & 20.40 & 5.23 \\ \midrule
      Predictive  & 16.77 & 0.71 &  7.93 & 8.13 \\ 
      StoRM       & 20.36 & 0.75 & 10.16 & 9.45 \\
      SB          & 15.44 & 0.67 &  5.08 & 9.69 \\ \bottomrule
    \end{tabular}
  }
\end{center}
\end{table}

%% file: tables/table_asr_models.tex
\begin{table*}
\caption{WER results obtained with different ASR models.}
\vspace{-0.5cm}
\label{table:asr_models}
\begin{center}
    \setlength\tabcolsep{1.0pt}
    \renewcommand{\arraystretch}{0.7}
    \resizebox{0.85\textwidth}{!}{
    \begin{tabular}{l @{\hskip 1em} *{12}{wc{4em}}}
      \toprule
      & \multicolumn{4}{c}{Fast Conformer Transducer Large} & \multicolumn{4}{c}{Parakeet RNNT 1.1B} & \multicolumn{4}{c}{Parakeet CTC 1.1B} \\
      \cmidrule(lr){2-5}\cmidrule(lr){6-9}\cmidrule(lr){10-13}
      Signal & WER/\% & INS/\% & DEL/\% & SUB/\% & WER/\% & INS/\% & DEL/\% & SUB/\% & WER/\% & INS/\% & DEL/\% & SUB/\% \\ 
      \midrule
      Unprocessed &  25.91 & 0.27 & 20.40 & 5.23 & 20.14 & 0.30 & 15.88 & 3.96 & 20.54 & 0.53 & 12.39 & 7.63 \\ \midrule
      SB & 15.44 & 0.67 & 5.08 & 9.69 & 13.01& 0.57 & 4.71 & 7.73 & 15.15 & 0.64 & 4.41 & 10.09 \\
      \bottomrule
    \end{tabular}
    }
\end{center}
\end{table*}

%% file: figures/number_steps.tex
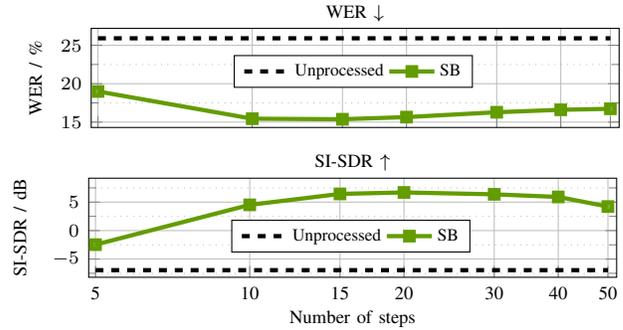
\begin{figure}
\centering
    \pgfplotstableread[col sep=comma]{tables/num_sampling_steps.csv}\results
    \begin{tikzpicture}
      \begin{semilogxaxis}[
          width=8.6cm,
          height=2.9cm,
          title=WER~$\downarrow$,
          title style={yshift=-1.5ex},
          xlabel style={yshift=1.3ex},
          xmajorgrids=true,
          minor x tick num=9,
          ylabel={WER / \%},
          ylabel style={yshift=-3.1ex},
          ymajorgrids=true,
          yminorgrids=true,
          minor y tick num=1,
          minor grid style={dotted},
          xtick={5,10,15,20,30,40,50},
          xticklabels={},
          xtick pos=bottom,
          legend style={legend columns=-1, nodes={scale=0.65}, draw=black, at={(0.27,0.40)}, anchor=south west},
          title style={font=\scriptsize},
          label style={font=\scriptsize},
          tick label style={font=\scriptsize},
          xtick align=inside,
          xmin=4.85,
          xmax=52,
          ymin=14.5,
          ymax=27,
          enlarge x limits = {abs=1},
          enlarge y limits = {abs=0.2}
        ]
        \addplot[draw=black, dashed, ultra thick] table[x=num_steps, y=unproc_wer]{\results};
        \addplot[draw={rgb:red,118;green,185;blue,0}, fill opacity=0.8, ultra thick, mark=square*, mark options={scale=0.8, fill={rgb:red,118;green,185;blue,0}}] table[x=num_steps, y=SB_wer]{\results};
        \legend{Unprocessed, SB}
      \end{semilogxaxis} 
    \end{tikzpicture}
    \begin{tikzpicture}
      \begin{semilogxaxis}[
          width=8.6cm,
          height=2.9cm,
          title=SI-SDR~$\uparrow$,
          title style={yshift=-1.5ex},
          xlabel=Number of steps,
          xlabel style={yshift=1.3ex},
          xmajorgrids=true,
          minor x tick num=9,
          ylabel={SI-SDR / dB},
          ylabel style={yshift=-2.0ex},
          ymajorgrids=true,
          yminorgrids=true,
          minor y tick num=1,
          minor grid style={dotted},
          xtick={5,10,15, 20,30,40,50},
          xticklabels={5,10,15,20,30,40,50},
          xtick pos=bottom,
          legend style={legend columns=-1, nodes={scale=0.65}, draw=black, at={(0.27,0.25)}, anchor=south west},
          title style={font=\scriptsize},
          label style={font=\scriptsize},
          tick label style={font=\scriptsize},
          xtick align=inside,
          xmin=4.85,
          xmax=52,
          ymin=-8,
          ymax=9,
          enlarge x limits = {abs=1},
          enlarge y limits = {abs=0.2}
        ]
        \addplot[draw=black, dashed, ultra thick] table[x=num_steps, y=unproc_sisdr]{\results};
        \addplot[draw={rgb:red,118;green,185;blue,0}, fill opacity=0.8, ultra thick, mark=square*, mark options={scale=0.8, fill={rgb:red,118;green,185;blue,0}}] table[x=num_steps, y=SB_sisdr]{\results};
        \legend{Unprocessed, SB}
      \end{semilogxaxis}
    \end{tikzpicture}
\vspace{-0.4cm}
\caption{WER and SI-SDR for the SB model with ODE sampler (configuration 4 in Table~\ref{table:models}) and different number of sampling steps.}
\label{fig:num_steps}
\end{figure}

%% file: figures/noise_levels_pred.tex
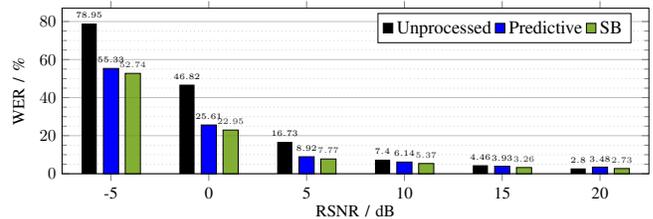
\begin{figure}
\centering
\pgfplotstableread[col sep=comma]{tables/noise_levels_pred.csv}\mydata

\begin{tikzpicture}[scale=0.63] 
    \begin{axis}[
            ybar=4pt,
            bar width=0.32cm,
            width=0.77\textwidth, 
            height=0.28\textwidth, 
            symbolic x coords={-5,0,5,10,15,20},
            xtick=data,
            nodes near coords,
            nodes near coords align={vertical},
            nodes near coords style={font=\tiny},
            ymajorgrids=true,
            yminorgrids=true,
            minor y tick num=3,
            minor grid style={dotted},
            ymin=0,ymax=87,
            ylabel={WER / \%},
            ylabel style={yshift=-2ex},
            legend pos=north east,
            legend style={legend columns=-1, nodes={scale=1.1}},
            legend image code/.code={\draw [#1] (0cm,-0.1cm) rectangle (0.2cm,0.1cm); },
            xlabel={RSNR / dB}
        ]
        \addplot[draw=none,fill=black] table[x=level,y=wer_unproc]{\mydata};
        \addplot[fill=blue] table[x=level,y=wer_pred]{\mydata};
        \addplot[fill={rgb:red,118;green,185;blue,0}, fill opacity=0.8] table[x=level,y=wer_sb]{\mydata};
        \legend{Unprocessed, Predictive, SB}
    \end{axis}
\end{tikzpicture}
\vspace{-0.3cm}
\caption{ASR performance in terms of WER vs. RSNR for the best SB model from Table~\ref{table:models} using ten sampling steps.}
\label{fig:noise_levels}
\end{figure}